\documentclass[useAMS,fleqn,usenatbib]{mnras}
\usepackage[T1]{fontenc}
\usepackage{ae,aecompl}
\usepackage{bethmacros}
\usepackage{graphicx}
\usepackage{amssymb}
\usepackage{amsmath}

\def\spose#1{\hbox to 0pt{#1\hss}}
\def\lta{\mathrel{\spose{\lower 3pt\hbox{$\mathchar"218$}}
     \raise 2.0pt\hbox{$\mathchar"13C$}}}
\def\gta{\mathrel{\spose{\lower 3pt\hbox{$\mathchar"218$}}
     \raise 2.0pt\hbox{$\mathchar"13E$}}}

\title[Limits of SN Feedback]{Cosmological Galaxy Evolution with Superbubble
Feedback II:  The Limits of Supernovae}

\author[Keller et al.]{B.\,W. Keller$^{1}$\thanks{Email: kellerbw `at' mcmaster.ca},  J. Wadsley$^{1}$, H. M. P. Couchman$^1$
\vspace*{6pt}\\
$^1$Department of Physics and Astronomy, McMaster University, Hamilton, Ontario, L8S 4M1, Canada
}

\begin{document}
\maketitle
\label{firstpage}
\begin{abstract}

  We explore when supernovae can (and cannot) regulate the star formation and bulge growth in galaxies based on a sample of 18 simulated galaxies.
  The simulations include key physics such as evaporation and conduction,
    neglected in prior work, and required to correctly model superbubbles resulting from stellar feedback.
    We show that for galaxies with virial masses $>10^{12}\;\Msun$, supernovae
    alone cannot prevent excessive star formation.  This failure
    occurs due to a shutdown of galactic winds, with wind mass loadings falling
    from $\eta\sim10$ to $\eta<1$.  In more massive systems, this transfer of baryons to
    the circumgalactic medium falters earlier on and the galaxies diverge significantly from observed galaxy scaling relations and morphologies.
    The decreasing efficiency is
    simply due to a deepening potential well preventing gas escape.  This implies that non-supernova feedback mechanisms
    must become dominant for galaxies with stellar masses greater than
    $\sim4\times10^{10}\;\Msun$.  The runaway growth of the central stellar
    bulge, strongly linked to black hole growth, suggests that feedback from active galactic
    nuclei is the probable mechanism.
    Below this mass, supernovae alone are able to produce a realistic stellar
    mass fraction, star formation history and disc morphology.
\end{abstract}


\section{Introduction}
Stellar feedback plays many roles in galaxies.  Simply regulating the star
formation rate within the interstellar medium (ISM) is not enough to produce a
realistic galaxy population.  The reduced baryon fractions seen in galaxies requires the
expulsion of that gas \citep{Mathews1971,Larson1974}.  Matching observed
scaling relations, cosmic star
formation histories, the stellar mass-metallicity regulation, and the stellar mass
function all require the ejection of gas from a galaxy in large-scale outflows
\citep{Dekel1986,Finlator2008,Erb2008,Peeples2011,Marasco2012}.  Both abundance
matching \citep{Behroozi2013,Moster2013} and gravitational lensing studies
\citep{Hudson2015} find that no galaxies convert more than $\sim25\%$ of their
inital baryonic mass into stars.  The most straight forward explanation for this is the
ejection of a significant fraction of the baryons from the galaxy disc.  

The observational evidence of galactic outflows is impossible to ignore (a
review of both the theory and observations of galactic winds can be found in
\citealt{Veilleux2005}).  From intergalactic metals seen in quasar absorption
lines \citep{Songaila1996,Dave1998,Weiner2009}, to the detection of neutral
\citep{Kunth1998,Morganti2003}, ionized \citep{Heckman1987,Martin2012}, and
molecular \citep{Stark1984} outflowing gas with outflow velocities $\gg
100\;\kms$ \citep{Leroy2015}, we can see that galaxies do not merely
accrete gas: they eject it as well (a detailed study of gas inflows and outflows
can be found in \citet{Woods2014}). In our own Milky Way, X-ray observations
have detected $\sim6\times10^{10}M_\odot$ in the $\sim10^6\K$ circum-galactic medium
(CGM) \citep{Gupta2012}, comparable in mass to total baryonic mass of the galactic disc.

In \citet{Keller2015}, we showed that a correct treatment of superbubbles driven
by supernovae (SN) directly generates strong galactic winds without requiring extra
wind models.  This was the first time the complete physics of thermal conduction
and evaporation required to model superbubbles (see \citealt{Keller2014}) was
employed in galaxy formation simulations.  This removes the need for cooling
shutoffs, hydrodynamic decoupling, or other purely numerical boosts to the
feedback effectiveness and the associated free parameters.  These first
principles outflows give a stellar mass evolution that matches the abundance
matching results \citep{Behroozi2013}.  By removing preferentially low-angular
momentum gas, these outflows can prevent the growth of a large bulge, producing
bulgeless discs like those that are seen in the nearby universe.

Prior work \citep{Hopkins2014,Agertz2015,Stinson2013} has examined the role of
outflows in enabling galaxies to regulate their baryon content and match
observed relations as a function of their mass.  A common outcome is that
stellar feedbacks are insufficient for higher mass galaxies.  Typically, these
works relied on multiple feedbacks and detailed subgrid models with associated
free parameters.  These additional feedbacks are discussed in more detail in section 2.  They
did not however, include necessary physics to accurately model superbubbles.
Thus the question what supernovae driven outflows can do has yet to be fully
answered.   For example, \citet{Hopkins2014} has argued that supernovae alone
cannot regulate the stellar content of intermediate mass galaxies, let alone the
more massive ones.  

The primary goal of this work is to examine galaxies and their outflows via a suite of
simulated field $L*$ galaxies, the McMaster Unbiased Galaxy Simulations 2
(MUGS2).  We begin with an overview of stellar feedback and galactic outflows
in section 2.  We examine which forms of feedback are most likely to
contribution outflows and also discuss alternatives to supernovae.  In section
3,  we describe the MUGS2 sample and simulation methods.  We then examine the
evolution and final state of the galaxy sample in section 4.   As in other work,
we find that galaxies more massive than $10^{12}\;\Msun$ deviate from observed
relations.  In our case, this is where supernova-driven winds begin to fail.  We
find a universal relation between the halo/disc mass and outflow mass loadings.
Finally, in section 5, we discuss how stellar mass regulation through supernovae
fails, how this manifests in the galaxy properties and how this provides strong
clues that Active Galactic Nuclei (AGN) must take over the regulation.

\section{What Launches Galactic Outflows?}
The question of what ultimately powers galactic winds has been debated for over
half a century, since the first outflows were discovered in M82
\citep{Lynds1963}.  Unfortunately, neither (AGN) nor
stellar processes have unambiguous observational signatures in the outflows they
produce \citep{Veilleux2005}.  To complicate matters further, stellar feedback
comes in multiple forms.  Disentangling these adds to the uncertainty of how
galactic outflows are actually generated.  Massive stars deposit energy and
momentum in the ISM through ionizing radiation, radiation pressure on dust
grains, stellar winds, and ultimately explode as supernova.  Each of these
processes, in principle, has sufficient energy to drive a galactic outflow.
Radiation, in particular, has orders of magnitude more energy available than the
others \citep{Leitherer1999}.  Driving effective galactic winds requires strong
coupling to the ISM gas and limited cooling losses so that the terminal velocies
are high relative to the escape velocity.  We examine the feedback processes
individually below.

\subsection{Stellar Feedback}
\subsubsection{Ultraviolet Radiation \& HII Regions}
The characteristic temperature of gas photoionized by UV radiation is
$\sim10^{4}\K$ \citet{Krumholz2009}, far lower than the virial temperature
of all but the smallest galaxies.  For galaxies with halo masses below
$10^9\Msun$, photoheating and photoionization by UV radiation strong limits star
formation \citep{Efstathiou1992}.  The characteristic sound speed for gas in HII
regions is $\sim10\;\rm{km/s}$ is similar to characteristic turbulent velocities
and escape speeds in molecular clouds.  Even large HII regions, such as 30
Doradus, have been observed to have expansion rates of only $25\;\rm{km/s}$
\citep{Chu1994}, significantly less than what is needed to drive even a weak
galactic fountain.  Thus it is doubtful that UV radiation plays much role in
launching galactic-scale outflows.  

Simulations by \citet{Dale2012} have shown that HII regions alone are unable to
effectively remove gas from their birth clouds, let alone their galaxies but can
alter their structure.  Thus UV can play a local role in simulations that
resolve molecular clouds.

\subsubsection{Radiation Pressure}
Radiation pressure on the dust grains in galaxies is a potential driver for
galactic outflows.  \citet{Murray2011} showed that, assuming spherical symmetry,
the most massive star clusters can drive reasonably fast ($v\sim$ hundreds of
km/s) outflows.  \citet{Agertz2015} and \citet{Hopkins2014} also found that a
combination of SN, stellar winds, and radiation pressure produced a galaxy which
matched the abundance-matched stellar mass fractions of \citet{Behroozi2013}.
However, as \citet{Agertz2015} noted, and \citet{Roskar2014} examined in detail,
these simulated galaxies often displayed unrealistic morphologies, with much
thicker discs than would be expected for a normal Milky-Way like disc.
\citet{Roskar2014} found the addition of radiation pressure using a local `UV
escape probability' model for dust absorption, with a single parameter for the
IR dust opacity $\kappa_{IR}$ was able to reduce the stellar mass fraction of a
cosmological galaxy with mass comparable to the lighter members of our
well-regulated population.  However, this required large values of
$\kappa_{IR}$, such that the UV radiation coupled to the ISM so strongly that
the entire disc was completely disrupted, and the resulting galaxy was
completely spheroidal, with stellar scale heights above $3\;\kpc$.

The porosity of the ISM can significantly reduce the mean optical depth of the
galaxy, giving photons an escape valve to leave unimpeded \citep{Krumholz2013}.
The importance of radiation pressure remains an open question.

\subsubsection{Supernovae and Stellar Winds}
Based on energetics alone, it might seem that SN alone could eject gas from even
the most massive halos.  A Type II SN releases $\sim10^{51}\;\rm{erg}$ in an
ejecta of $\sim10\;\Msun$ \citep{Leitherer1999}. This results in a maximum
ejecta temperature of $\sim2\times10^8\K$.  This corresponds to the virial
temperature for halos with a mass of $\sim10^{15}\;\Msun$.  This fails to take
into account the mixing of SN ejecta with the surrounding ISM.   Additionally,
if SN ejecta left a galaxy without any mixing or entrainment of additional gas,
the largest wind mass loadings seen would be $\eta\sim0.1$.  Such winds would
remove metals from the ISM but have little effect on the total baryon content.
In order to moderate both star formation and bulge growth, mass loadings of
$>10$ are necessary.  Mixing in cooler ISM material reduces the effective wind
temperature to $\sim10^6\K$ if cooling losses are small.

While it is well known that individual supernovae experience strong cooling,
clustered star formation allows supernovae and stellar winds to combine into a
superbubble that retains 65 \% of its initial injected energy even after the
formation of cold shell which then breaks out of the ISM into the halo
\citep{MacLow1988}.  With low density channels up to 99 \% of the energy can
escape \citep{Rogers2013}.  An important function of the early stellar feedbacks
discussed earlier is to clear dense gas around the star cluster to enable the
superbubble to escape.  However, as discussed previously, this is only valid in
a simulation if dense gas is resolved.  Thus superbubble feedback can be much
more efficient that supernova feedback.  The physics of evaporation also leads
to specific predicitions with respect to mass loading and outflow temperatures
of order $\sim 10^6\K$ \citep{Keller2014}.

$\sim10^6\K$ is the virial temperature of halos with masses of a few
$10^{12}\;\Msun$.  If the wind fluid is cooler than the virial temperature, it
will not be rise buoyantly out of the disc into the circum-galactic medium
(CGM).  This suggests SN driven winds become less effective, and may fail to
launch altogether somewhere in the mass range of $\sim10^{12}\;\Msun$. The
existence of a peak in the star formation efficiency at this same mass, as seen
in \citet{Behroozi2013} is strong evidence that this indeed be occurring in
nature, and that some other feedback mechanism begins to dominate at higher
masses. Finding out the if, when and how of this transition is important if we
want to know how larger galaxies quench their star formation, and if SN alone
can explain the low star formation efficiency in galaxies below this mass.

\subsection{Other Feedback Mechanisms}
\subsubsection{AGN}
The primary non-stellar energy source available for driving galactic outflows is
feedback from the growth of super-massive black holes (SMBHs).  The Milky Way's
own SMBH has a mass of $M_{\bullet}\sim4\times10^{6}\;\Msun$ \citep{Meyer2012}.
The energy released by it's formation is
$M_{\bullet}c^2\sim7\times10^{60}\;\rm{erg}$.  This is significantly greater
than the binding energy of the galactic halo ($\sim10^{59}\;\rm{erg}$).  If even
a small fraction of this energy couples to the ISM as it is released, it can
have a significant disruptive effect.  SMBHs are ubiquitous and larger ones are
strongly linked to the presence of massive bulges
\citep{Magorrian1998}.

Much effort has gone into developing sub-grid models for AGN feedback and SMBH
growth, and now AGN feedback is a major component of many large box simulations
(e.g., Illustris \citep{Sijacki2015}, EAGLE \citep{Crain2015}, Rhapsody-G
\citep{Hahn2015}, Horizon-AGN \citep{Dubois2014})

\subsubsection{Cosmic Rays}
Cosmic rays (CRs) have been been proposed as another potential engine for
driving outflows \citep{Ipavich1975,Breitschwerdt1991,Everett2008,Socrates2008}.
Cosmic rays could be directly linked to SN shocks or shocks within the ISM.
Cosmic rays may contain as much energy as the thermal and magnetic components of
a galaxy \citep{Zweibel1995}.

\citet{Jubelgas2008} found CR had little impact on higher mass galaxies.
\citet{Girichidis2015} showed that cosmic rays can launch winds with mass
loadings of order unity for gas surface densities comparable to the Milky Way.
\citet{Salem2014} and \cite{Booth2013} found CR driven outflows had even lower
mass loadings for Milky Way mass halos.  These results make it doubtful that
cosmic rays are centrally important for galactic outflows.

\section{Methods: MUGS2}
The simulations presented here are the new MUGS2 simulations.  The original MUGS
sample, presented in \citep{Stinson2010},  followed the evolution of isolated,
Milky-Way like disc galaxies using low temperature metal cooling, UV background
radiation, and stellar feedback. The new MUGS2 sample includes all of this, 
plus a number of improvements to the hydrodynamic method, along with
the new superbubble feedback model \citep{Keller2014}.  This has resulted in
significantly different evolution in the MUGS2 sample compared to the original
MUGS set, most notably greatly reduced star formation in nearly every galaxy
(the original MUGS sample greatly overproduced stars).

These simulations were run using the modern smoothed particle hydrodynamics
(SPH) code {\sc GASOLINE2}, as in \citet{Keller2015} and \citet{Keller2014}.
The changes in this new code include a sub-grid model for turbulent mixing of
metals and energy \citep{Shen2010}, and a modified pressure force form similar
to that proposed by \citet{Ritchie2001}, which is functionally equivalent to
\citet{Hopkins2013}.  Details of the star formation and gas physics model can be
found in \citet{Keller2014}.

\subsection{Simulation Initial Conditions}
We adopt the same initial conditions (ICs) that were used in the original
\citet{Stinson2010} MUGS sample.  These cosmological zoom-in ICs were selected
from a $50 h^{-1}\;\rm{Mpc}$ cube, run (with dark matter only) to $z=0$.  The
simulations use a {\sc WMAP3} $\Lambda\rm{CDM}$ cosmology, with $H_0 =
73\;\rm{km/s/Mpc}$, $\Omega_M = 0.24$, $\Omega_{bary} = 0.04$, $\Omega_\Lambda =
0.76$, and $\sigma_8 =0.76$.  Galaxies that had halo mass from
$5\times10^{11}\;\Msun$ to $2\times10^{12}\;\Msun$ were then selected from the
dark matter only run.  Isolated galaxies were then selected by choosing only
ones which had no neighbours in this mass range within $2.7\;\rm{Mpc}$.  This
resulted in a sample of 276 candidate halos, of which the final MUGS sample was
simply selected at random from this pool.  While only 9 of these were presented
in \citet{Stinson2010}, subsequent papers presented the remaining galaxies
\citep{Bailin2010,Nickerson2011,Nickerson2013}, and what we present here is the
full set of galaxies that were produced for MUGS.

Each of these simulations has a gas mass resolution of $M_{gas}=2.2\times10^5\;
\rm{M_{\odot}}$, and uses a gravitational softening length of $\epsilon=312.5\;
\rm{pc}$, and a minimum SPH smoothing length set to 1/4 of this.  The total
sample presented here consists 18 galaxies, with z=0 virial masses in a range
from $3.7\times10^{11}\;\Msun$ to $2.1\times10^{12}\;\Msun$.

\subsection{Star Formation}
We use a standard Schmidt-law star formation recipe, where the rate is set by
the freefall time of gas and an efficiency $c_*$: \begin{equation} \dot \rho_* =
\frac{c_*\rho}{t_{ff}} = c_*\sqrt{\frac{32G}{3\pi}}\rho^{3/2} \end{equation} We
used an efficiency of $c_*=0.05$, as was used for the original MUGS simulations
\citep{Stinson2010} and in \citet{Keller2015}.  Star formation is only allowed
to take place in gas which satisfies three criteria: it has a temperature below
$1.5\times10^4\K$, density above $9.6\;\rm{cm^{-3}}$ (the density where
gravitational softening becomes important), and is in a converging flow
($\nabla\cdot \vec v < 0$).  This density threshold is larger than the that used
in \citet{Stinson2010} (which used $0.1\;\rm{cm^{-3}}$, but identical to that
used in \citet{Stinson2013}.  We have used this larger threshold (as
\citet{Governato2010} recommends) to better capture the effects of clustered
star formation.  Aside from this higher star formation threshold, our star
formation recipe remains unchanged from the original MUGS sample.

\subsection{Superbubble Feedback}
\begin{table*}
\begin{tabular}{rrrrrrrrrr}
	\hline
	Galaxy & $M_{vir}$ & $\lambda'$ & $f_b$ & $z_{1/2}$ & $z_{lmm}$ &  $M_*$ & $M_{gas}$ & $M_{central}$ & $SFR_{z=0}$ \\
	ID&&&&&&  (masses & in & $10^{10}\;M_\odot$) & $(M_\odot\; yr^{-1})$ \\
	\hline
	\hline
	g7124 & 36.6 & 0.039 & 0.150 & 0.8 & 2.1 & 0.5 & 5.0 & 1.8 & 0.6 \\
	g5664 & 47.7 & 0.028 & 0.173 & 0.8 & 3.0 & 0.9 & 7.3 & 3.0 & 1.7 \\
	g8893 & 58.0 & 0.065 & 0.170 & 1.0 & 0.8 & 0.7 & 9.1 & 2.6 & 1.0 \\
	g1536 & 64.9 & 0.029 & 0.189 & 1.0 & 4.0 & 1.9 & 10.4 & 5.2 & 3.9 \\
	g21647 & 74.4 & 0.069 & 0.152 & 0.2 & 0.5 & 1.2 & 10.1 & 3.1 & 2.7 \\
	g422 & 76.2 & 0.033 & 0.183 & 0.8 & 0.7 & 1.5 & 12.4 & 5.1 & 3.2 \\
	g22437 & 85.2 & 0.013 & 0.192 & 0.7 & 2.6 & 9.0 & 7.3 & 11.2 & 17.9 \\
	g22795 & 85.2 & 0.009 & 0.178 & 1.1 & 3.8 & 10.6 & 4.6 & 11.7 & 6.2 \\
	g3021 & 97.8 & 0.040 & 0.192 & 0.6 & 1.7 & 3.6 & 15.1 & 9.9 & 15.7 \\
	g28547 & 98.5 & 0.106 & 0.186 & 0.4 & 0.1 & 1.6 & 16.7 & 4.5 & 2.7 \\
	g19195 & 101.6 & 0.039 & 0.162 & 0.7 & 0.7 & 7.1 & 9.3 & 9.7 & 20.3 \\
	g24334 & 102.2 & 0.052 & 0.174 & 0.5 & 1.6 & 2.6 & 15.3 & 7.2 & 6.8 \\
	g4720 & 102.5 & 0.015 & 0.192 & 0.8 & 1.8 & 14.2 & 5.5 & 14.4 & 8.3 \\
	g4145 & 119.5 & 0.033 & 0.193 & 0.9 & 1.4 & 15.0 & 8.1 & 16.9 & 20.9 \\
	g25271 & 125.5 & 0.016 & 0.187 & 1.1 & 4.0 & 15.6 & 7.9 & 17.0 & 9.7 \\
	g15784 & 131.2 & 0.037 & 0.186 & 1.3 & 7.3 & 13.0 & 11.4 & 17.7 & 9.6 \\
	g15807 & 203.2 & 0.026 & 0.191 & 1.0 & 2.5 & 21.4 & 17.5 & 25.4 & 11.9 \\
	g27491 & 214.7 & 0.039 & 0.184 & 0.7 & 1.1 & 18.8 & 20.8 & 26.6 & 17.6 \\
	\hline
\end{tabular}

\caption{Redshift 0 properties of the full MUGS2 sample.  All masses and
    particle counts are measured within a $R_{vir}$ sphere centered on the halo, except for
$M_{central}$, which is measured within a $0.1\;R_{vir}$ sphere.}
\label{z0_table}
\end{table*}
\begin{figure*}
    \includegraphics[width=\textwidth]{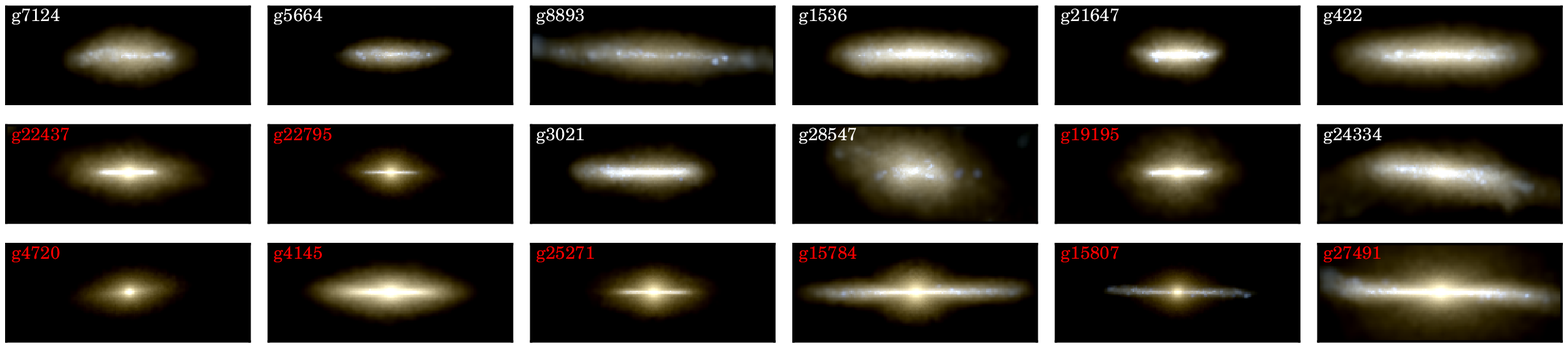}
    \includegraphics[width=\textwidth]{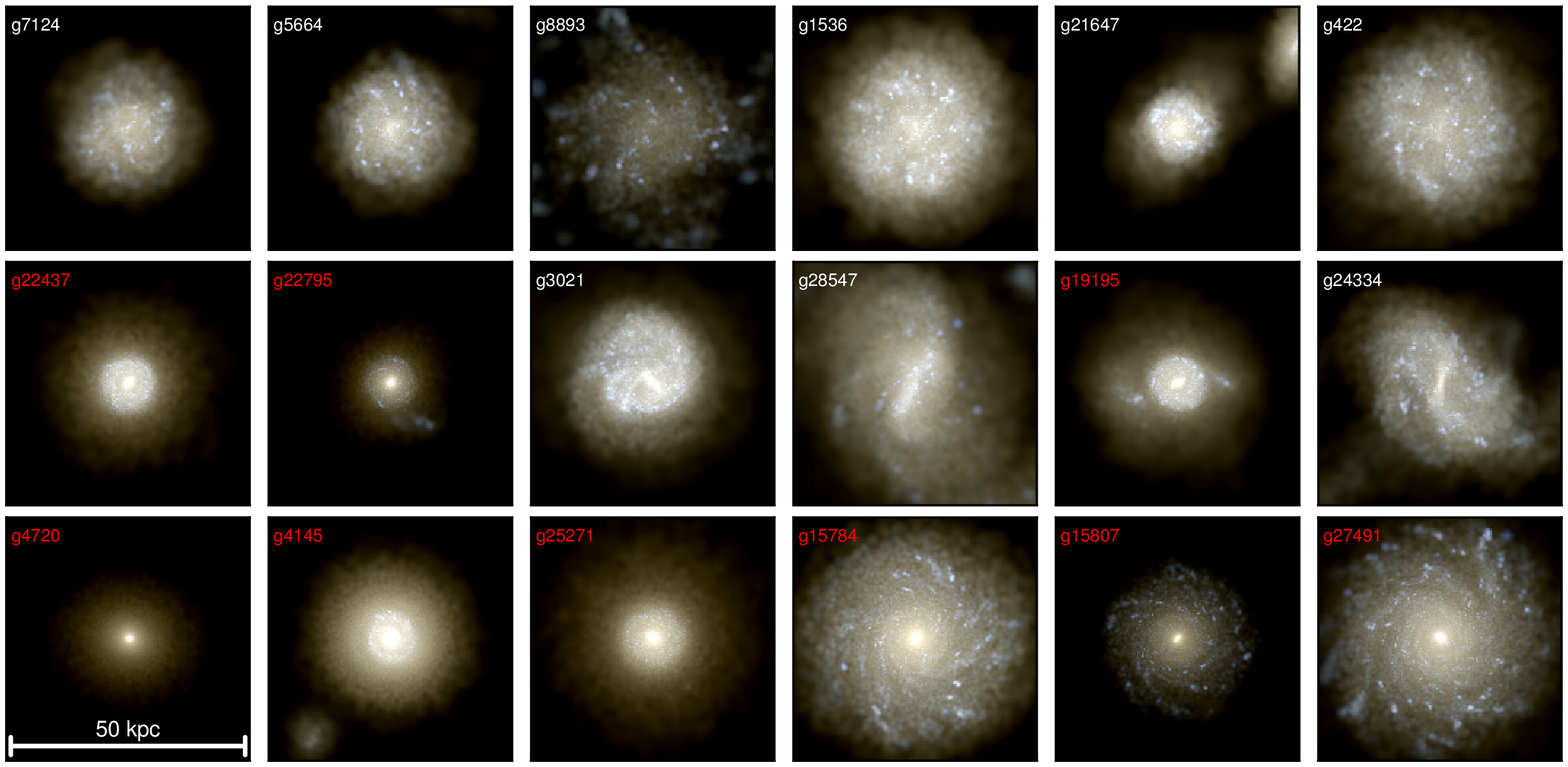}
    \caption{Mock optical stellar observations of each MUGS2 galaxy.  RGB
        channels are calculated using \citet{Marigo2008} stellar populations.
        The top 3 rows show the galaxies edge on, while the bottom 3 show the
        galaxies face on.  The galaxies are sorted in order of halo mass, and
        galaxies that overproduce stars are labelled with red.  These images do
    not include the effects of dust extinction}
    \label{stellar_image}
\end{figure*}

We use the superbubble feedback model described in \citet{Keller2014}.  We
briefly summarize the important details below.  The model deposits thermal
energy and mass into resolution elements in a brief multi-phase state.  These
particles each have separate specific energies and masses for the hot bubble and
cold ISM components, including the swept up shell.  This allows the method to
calculate a separate density, temperature, and cooling rate for each phase,
rather than using the average density and temperature of both phases.
Multiphase particles are also prevented from forming stars if the average
temperature of the two phases is above our temperature threshold for star
formation.  The amount of supernova feedback per unit stellar mass is determined
using a \citet{Chabrier2003} IMF.  

In \citet{Keller2014}, we demonstrated that superbubble feedback could produce
significantly more mass-loaded outflows and reduced star formation in
simulations of isolated disc galaxies, compared to the older \citep{Stinson2006}
blastwave model.  We also showed how the ISM phase behaviour in
superbubble-regulated galaxy behaved in a much more physically reasonable way
than in a galaxy regulated with single phases.  \citet{Keller2015} showed that
in a cosmological galaxy (one of the same galaxies presented here in the MUGS2
sample, g1536) superbubble feedback resulted in high mass loadings $\eta\sim10 $
for galactic outflows from $z\sim4-2$, which then fell to $\eta\sim1$ at low
redshift, producing a galaxy with a realistic star formation history and no
significant bulge component, without the need for additional feedback
mechanisms, or greater feedback energies than what are provided for simply by SN
alone.  We also showed that this comes without additional computational cost, or
the addition of any new free parameters/tuning.

With $\sim10^{51}\; \rm{erg}$ coupled to the ISM per supernova this gives
$\sim10^{49}\; \rm{erg\;M^{-1}_{\odot}}$ for the full stellar population.  This
is significantly less energy per mass of stars formed than models such as the
FIRE \citep{Hopkins2014} model, which included feedback from stellar winds and
radiation pressure, or the large volume simulations like EAGLE
\citep{Schaye2015} or Illustris \citep{Sijacki2015}, which required artificially
boosted feedback energies to match observed galaxy populations.

Aside from a different star formation density and temperature threshold, the
star formation and feedback recipes used here identical to those used in
\citet{Keller2014}, and are completely identical to those used in
\citet{Keller2015}.

\section{Results}

\begin{figure*}
    \includegraphics[width=\textwidth]{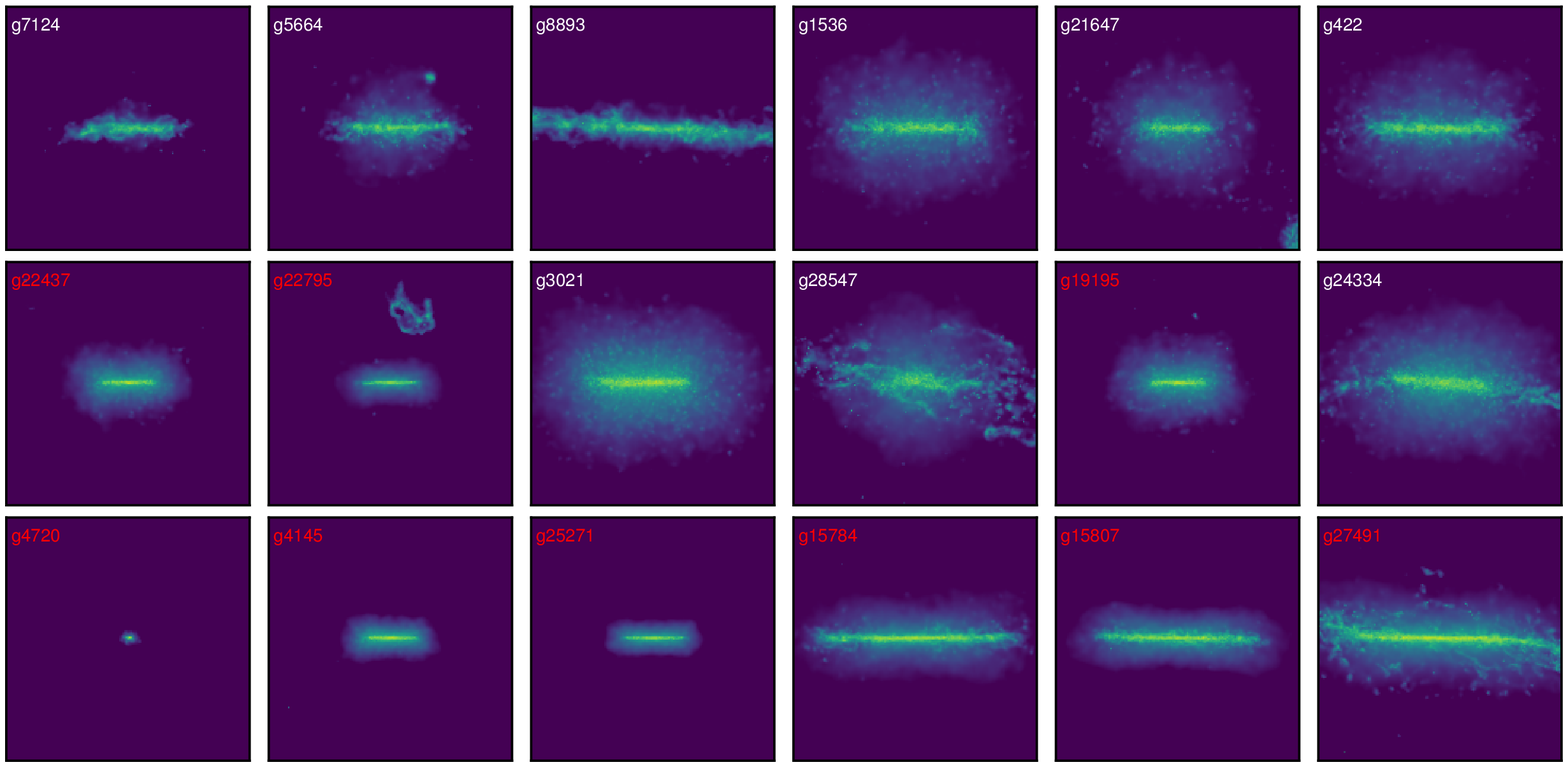}
    \includegraphics[width=\textwidth]{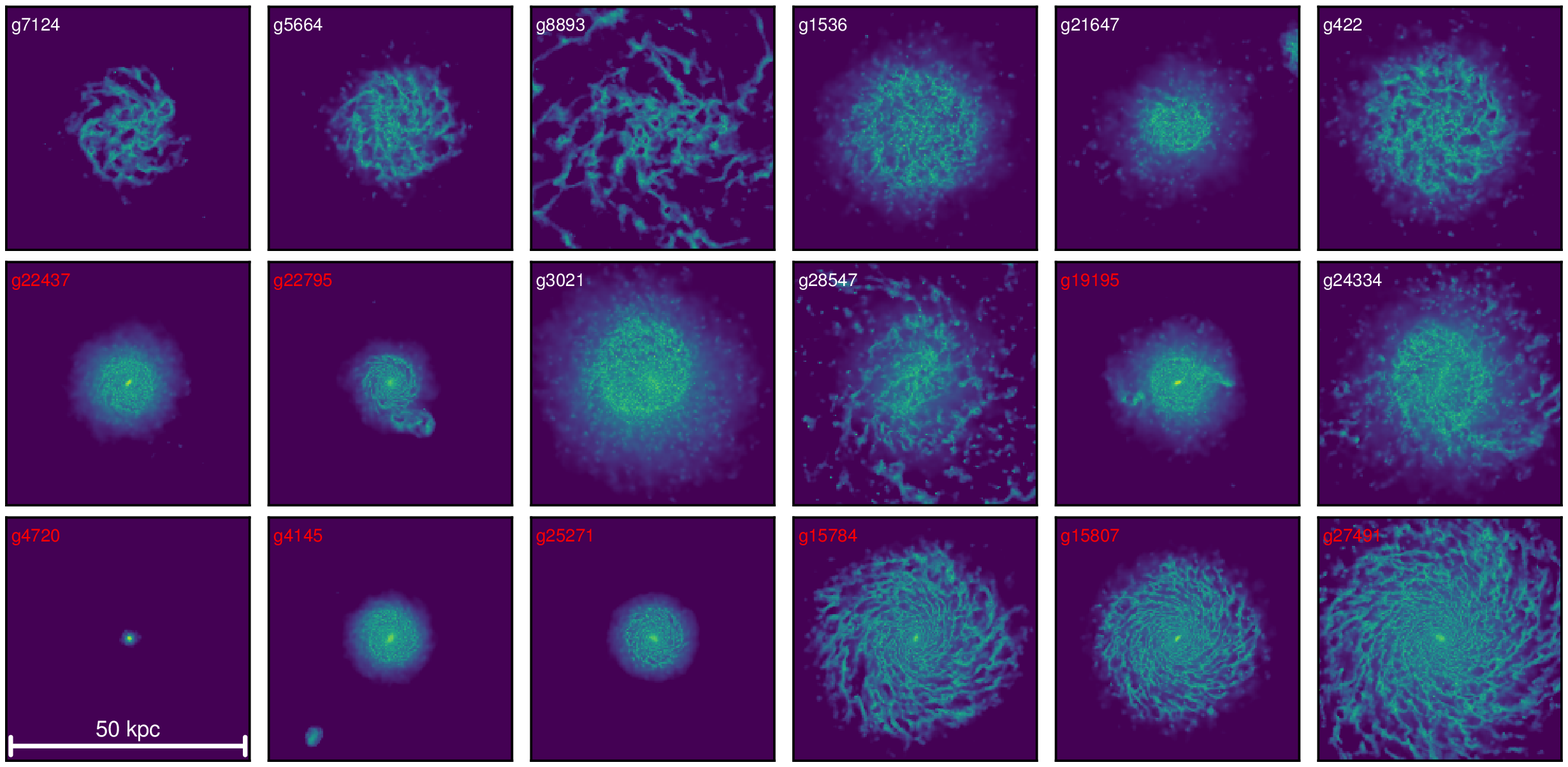}
    \caption{HI column density in each of the MUGS2 galaxies.  As in
    figure~\ref{stellar_image}, the top 3 rows show the galaxy edge on.}
    \label{gas_column}
\end{figure*}
\begin{figure}
    \includegraphics[width=\columnwidth]{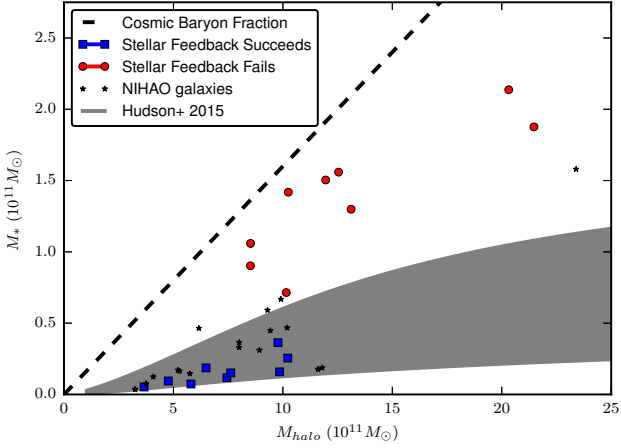}
    \caption{Stellar mass vs. halo mass at z=0.  The observed stellar mass to
        halo mass relation is from \citet{Hudson2015}'s weak lensing study of
        galaxies to z=0.8.  Galaxies that fall within this range are shown as
        blue squares, while galaxies above the curve (those which form too many
        stars) are shown as red circles.  For comparison, the black stars show
        galaxies from the NIHAO sample \citep{Wang2015}.}
    \label{SMHMR}
\end{figure}
\begin{figure}
    \includegraphics[width=\columnwidth]{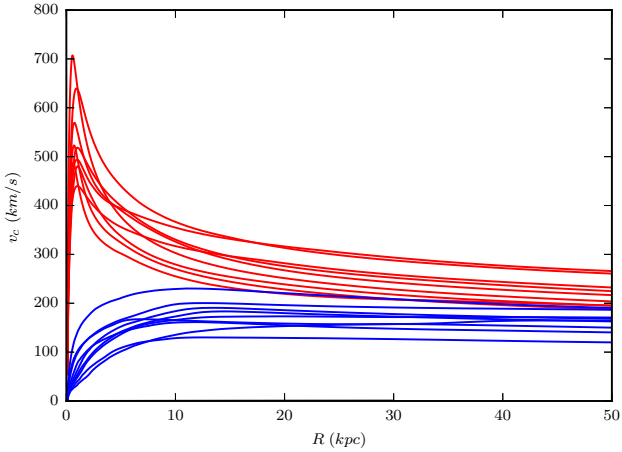}
    \caption{As is clear from the above rotation curves, galaxies which
    overproduce stars (shown in red) also have large central concentrations,
    giving steeply peaked rotation curves inconsistent with those seen in local $L*$
    galaxies.  Galaxies with well-regulated star formation have flat rotation
curves (shown in blue).}
    \label{rotation_curve}
\end{figure}
\begin{figure}
    \includegraphics[width=\columnwidth]{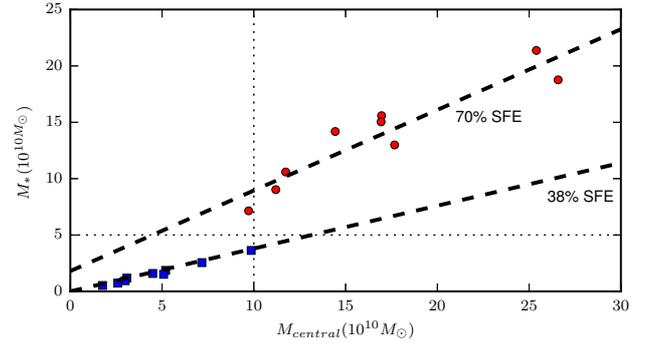}
    \caption{The stellar mass and the central baryonic mass show a tight
        correlation.  For the well-regulated population, roughly $40\%$ of the
        central baryons are stars, while for the unregulated population has
        converted $\sim70\%$ of the central baryon mass into stars.}
    \label{stellar_central}
\end{figure}
\begin{figure}
    \includegraphics[width=\columnwidth]{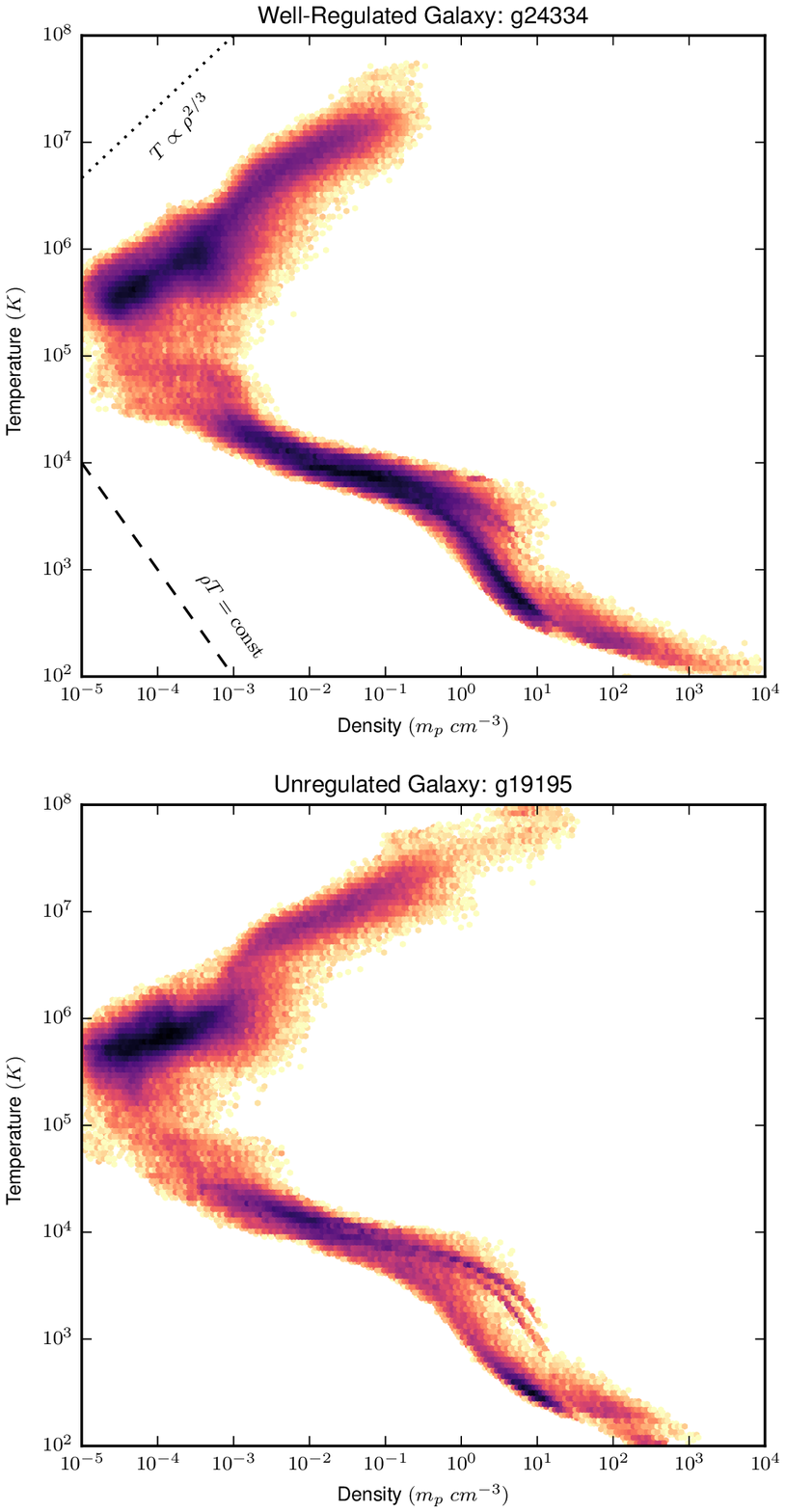}
    \caption{Gas phase diagrams for a well-regulated galaxy (g24334) and an
        unregulated galaxy (g19195).  Color shows the amount of mass at a given
        $\rho-T$ point.  The top phase diagram shows the characteristic slope
        for adiabatic (dashed line) and isobaric (dotted line) processes.  As
        can be seen, both galaxies are qualitatively the same.  Both show the
        equilibrium cooling curve below $10^4\K$, and both show the adiabatic
        evolution of superbubble-heated gas as it leaves the disc.}
    \label{phase}
\end{figure}

\subsection{Redshift Zero Properties}

The general properties of each galaxy at z=0 can be found in
table~\ref{z0_table}.  $\lambda'$ is the dimensionless spin parameter of the
halo, defined by \citet{Bullock2001} as $\lambda'=J/\sqrt{2GM_{vir}^3R_{vir}}$.
$f_b$ is the baryon fraction of the halo.  $z_{1/2}$ and $z_{lmm}$ are the
redshifts at which the halo reaches half of its final mass, and the redshift of
last major merger respectively.  Major mergers are defined to match
\citet{Stinson2010}, as a merger with a halo containing at least $1/3$ the
stellar mass of the main progenitor.  $M_{vir}$, $M_*$, and $M_{gas}$ are the
masses of the full halo, the stellar component, and the gas component
respectively within the virial radius.  $\rm{SFR_{z=0}}$ is the redshift 0 star
formation rate (SFR), averaged over the previous $100 Myr$.  The virial radius
is defined as the radius around the halo such that the enclosed density is 200
times the critical density ($\rho=200\rho_{crit}$).  While we used the same
definitions for quantities reported in \citet{Stinson2010}, every value reported
here is derived from the new MUGS2 simulations.  We also define a {\it central}
region of the halo, which contains the disc and the majority of the stars within
the halo.  The baryon mass within this region is given as $M_{central}$  This
central region is simply a sphere of radius $0.1R_{vir}$.

Mock stellar observations and HI column images of these galaxies can be seen in
figures~\ref{stellar_image} and~\ref{gas_column}.  The varied merger history of
these galaxies is evident in these images: companions can be seen in 3 galaxies
(g21647, g22795, g4145), tidal tails are evident in another  (g19195), and
strong bars exist in another 2 (g28547, g24334).  The HI column density shown in
figure~\ref{gas_column} shows, as was seen in \citet{Keller2015}, large
quantities of extraplanar HI gas, driven up by outflows from the galaxy disc.
The labels on each of these images are coloured red if they are in the
``unregulated'' population discussed in the next section.

\subsection{Stellar Mass Runaway}
As can be seen in figure~\ref{SMHMR}, roughly half of these galaxies fall within
the expected $z=0$ stellar mass to halo mass relation (SMHMR).  The grey band
shown in the figure is the $2\sigma$ confidence region in the observed SMHMR
from \citet{Hudson2015}, which used the CFHTLens weak lensing survey to produce
a fully-observational SMHMR, without the uncertainties of the using dark
matter-only simulations that are needed for abundance matching techniques (the
\citet{Hudson2015} SMHMR is consistent with past abundance matching estimates
such as \citet{Behroozi2013} and \citet{Moster2013}).  For the rest of this
paper, we will refer to the galaxies that fall within the $2\sigma$ confidence
interval for the observed SMHMR (i.e. galaxies where stellar feedback prevents
overproduction of stars) as {\it well-regulated}, and the galaxies where stellar
feedback fails to produce the correct SMHMR as {\it unregulated}.  No other
criterion is used for determining which population a galaxy falls into. In
addition to overproducing stars, these unregulated galaxies are qualitatively
redder and more bulge dominated, as can be seen in the mock stellar images of
figure~\ref{stellar_image}.  There appears to be no significant difference in
terms of merger activity between the two populations, strongly suggesting that
the failure of stellar feedback to regulate the galaxies is not simply a matter
of recent or violent merging.  The rotation curves in
figure~\ref{rotation_curve} also show the distinct signature of massive bulges
formed by catastrophic angular momentum loss in the unregulated population.
\citet{vanDenBosch2001} showed that this can arise when gas simply traces the
dark matter distribution, without redistribution or ejection by feedback.  The
strong peaks (as high as $700\;km/s$ for g4720) come about from the central
concentration of baryons in the galaxy bulge.  Without exception, each of the
well-regulated galaxies shows a flat rotation curve, with no evidence of a
significant bulge component.  This matches the qualitative morphology seen in
figure~\ref{stellar_image}.

Figure~\ref{stellar_central} shows that for the well-regulated population, the
stellar mass and central baryonic mass follow an extremely tight linear
relation, with a mean star formation efficiency (simply defined here as the
fraction of central baryons that are in stars) of $38\pm2\%$ over a Hubble time.
For the galaxies that fail to self-regulate, they exhibit a total star formation
efficiency of $70\pm10\%$, converting the majority of the baryons that collapse
onto their disc into stars.  Interestingly, the two populations can be divided
cleanly along the $M_{central}=10^{11}\;\Msun$ or the
$M_*=5\times10^{10}\;\Msun$ axis.  This is clear evidence that what explains
the dichotomy here must involve the accretion (or the failure to remove!)
baryons from the central region of the halo, where the disc resides and stars
are formed.

The gas phase diagram of galaxies in the two populations is remarkably similar,
as can be seen in figure~\ref{phase}.  For a representative pair from each
population, of nearly equal halo mass, gas follows essentially the same
evolutionary path, characteristic of a superbubble-regulated ISM, as was shown in
\citet{Keller2015}.  Gas accretes from the halo,
building up a warm ISM at $T\sim10^4\K$.  Where that gas reaches densities of
$\sim 1\;$cm$^{-3}$, it begins to cool quickly and form stars.  Those stars begin to
explode as SN, and a hot superbubble is formed, with $T>10^7\K$.  As the bubble
grows, it's temperature falls as it both expands adiabatically and evaporates
the cold shell surrounding it.  This hot, buoyant gas leaves the disc, rising
through the CGM and cooling adiabatically as it goes.  What is remarkable here
is just how similar the phase diagrams of these two galaxies are.  The slight
differences are what we would expect from table~\ref{z0_table}.  g19195 has
$\sim50\%$ less gas than g24334, and more than 3 times the SFR at $z=0$.  Thus,
there is less total material, especially in the warm \& cold phases, and
slightly more of the very hot ($T>5\times10^7\K$) gas (the interiors of very young
superbubbles) as a consequence of the higher SFR.  What is important to take
away from these phase diagrams is that a well-regulated and unregulated galaxy
do not have any real difference in the phase behaviour of their gas.  There is
no runaway cooling, or failure of SN to generate hot gas:  what causes the
unregulated population to runaway is purely an effect of the outflow
effectiveness, which is set by the depth of the potential well.

\subsection{Time Evolution} 
\begin{figure}
    \includegraphics[width=\columnwidth]{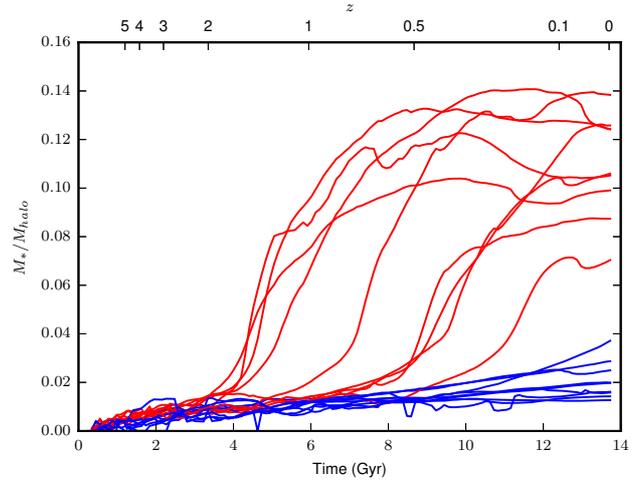}
    \caption{Stellar mass fraction vs. time.  As is clear, for the unregulated
        population the buildup of stellar mass above $\sim3\%$ of the halo mass
        happens very rapidly, often on timescales of $\sim1\;Gyr$.  The runaway
        in the unregulated population stops simply when the disc has been
        sufficiently depleted of gas that little remains to form new stars.}
    \label{stellar_fraction_time}
\end{figure}
\begin{figure}
    \includegraphics[width=\columnwidth]{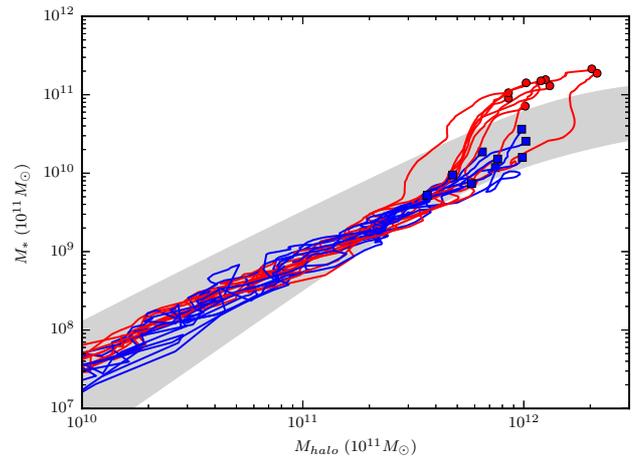}
    \caption{Stellar mass vs. halo mass, as shown in figure~\ref{SMHMR}, but
        with trails added to show time evolution.  For halo masses below
        $\sim\rm{a\;few}\times10^{11}$, the evolution of the regulated and
        unregulated populations are indistinguishable.  The grey bar shows the
        $z=0$ SMHMR, as in figure~\ref{SMHMR}}
    \label{SMHMR_time}
\end{figure}
\begin{figure}
    \includegraphics[width=\columnwidth]{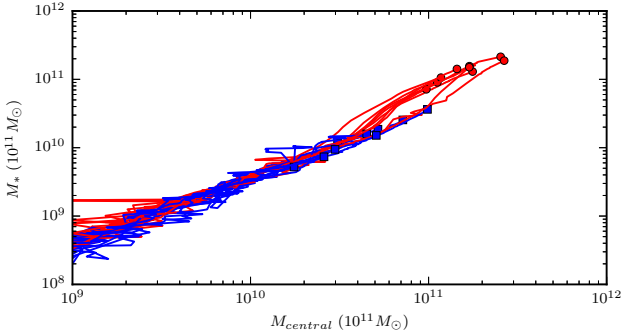}
    \caption{Stellar mass vs. central baryon mass. The stellar mass here can be
    seen to vary much more smoothly over the evolution, without the slight
    ``jump'' seen in figure~\ref{stellar_fraction_time}.}
    \label{stellar_central_time}
\end{figure}

The obvious question that arises from the presence of these two (well-regulated
vs.  unregulated) populations at $z=0$ is whether they are distinct through
their entire evolution, and if not, why/how do they diverge?  Previous work
suggests that a characteristic (halo or stellar) mass exists above which AGN
feedback is needed.  There is much uncertainty, however, about where this mass
exactly is, and how the transition from SN to AGN regulation actually works. We
should thus expect galaxies to begin diverging once this characteristic mass is
exceeded.  As figure~\ref{stellar_fraction_time} shows, the stellar mass
fraction for the two populations appear to follow similar evolutionary tracks
for some time (until past $z=0.5$ in the case of g19195, the galaxy which
diverges the latest).  However, once a galaxy begins to overproduce stars, it
appears to do so in a runaway, doubling or even tripling its stellar mass
fraction in less than a Gyr.

The similar early star formation history of the two populations is even more
clear if we look at the SMHMR's time evolution in figure~\ref{SMHMR_time}.  Here
we can see, that for galaxies with halo masses below a few $10^{11}\;\Msun$,
the stellar mass falls well within the expected SMHMR distribution, and the two
populations are indistinguishable.  The mass at which the populations begin to
diverge ( $M_*\sim10^{10}\;\Msun$ and $M_{vir}\sim3 -
5\times10^{11}\;\Msun$) are quite close to the expected transition range
from \citet{Shankar2006} ( $M_*\sim1.2\times10^{10}\;\Msun$ and
$M_{vir}\sim3\times10^{11}\;\Msun$).

The relatively smooth change in the $M_*-M_{central}$ relation, shown in
figure~\ref{stellar_central_time}, shows that it is in fact the central baryon
mass that is more tightly correlated with the stellar mass than the total halo
mass (in galaxies regulated by SN feedback alone). In fact, it is likely that
some of the more massive galaxies in the well-regulated sample are on their way
to failing, but simply have not yet had enough time by $z=0$ to diverge
significantly from the observed SMHMR.  Once more than $\sim10^{11}\Msun$ of
baryons have accreted to the disc of a galaxy, SN alone appear to be unable to
halt the growth of stellar mass.  At this point, star formation efficiency of
the central disc begins to increase, from $\sim40\%$ to $\sim70\%$, resulting in
the differences seen in the two populations at $z=0$.  This accretion crisis, as
it continues to smaller scales, also explains the rotation profiles of the
unregulated population.  The overcollapse that results in runaway starformation
also builds a massive bulge, producing the peaked rotation curves seen in
figure~\ref{rotation_curve}.

\subsection{Galactic Outflows}
\begin{figure}
    \includegraphics[width=\columnwidth]{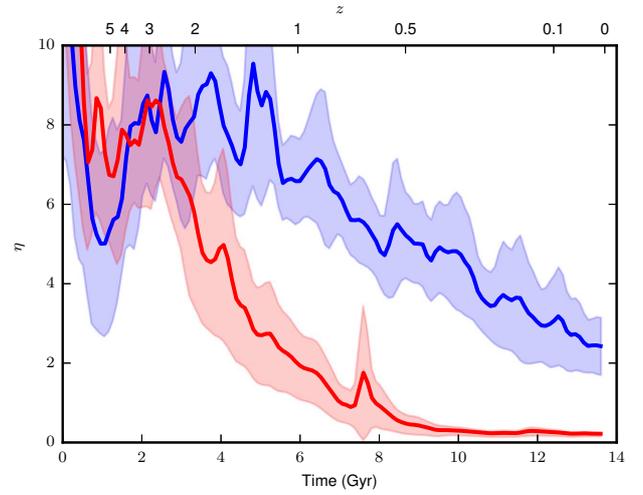}
    \caption{Outflow mass loading decreases as galaxies grow over time, as
    was previously shown by \citet{Keller2015}.  Here we see, when examining the
    mean mass loadings for the two populations (the well-regulated galaxies,
    where stellar feedback succeeds, and the unregulated galaxies, where it
    fails), that the unregulated galaxies have their mass loadings decrease
    sooner, and to lower values, than the well-regulated population.  The error
bars here are the $1\sigma$ scatter in each population.}
    \label{massloading_time}
\end{figure}
\begin{figure}
    \includegraphics[width=\columnwidth]{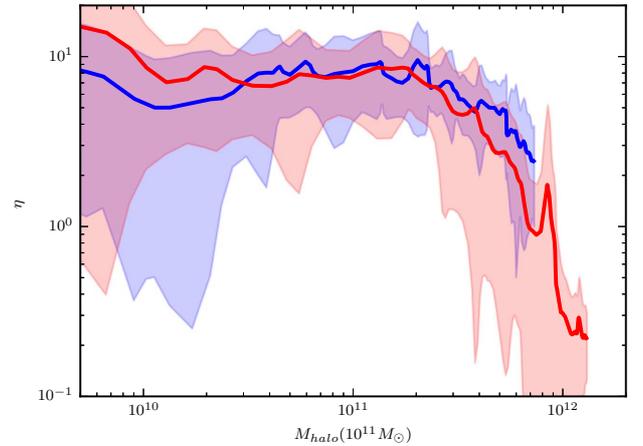}
    \caption{Unlike in the previous figure, if we look at mass loading as
    a function of halo mass, we instead see no significant difference between
    the two populations.  Instead, both follow a similar trajectory, and the halos which
    fail to self-regulate do so simply because they reach a higher mass earlier.}
    \label{massloading_halo}
\end{figure}

A key metric for both observational and simulation studies of galactic outflows
is the mass loading of these outflows.  This scaling is usually written as a
power-law scaling between mass loading $\eta$ and the circular velocity of the
halo or halo mass \citep{Murray2005,Peeples2011}:
\begin{equation}
    \eta \propto v_c^{-\alpha} = (GM_{halo}/R_{vir})^{-\alpha/2} \propto
    M_{halo}^{-\alpha/3}
\end{equation}
The normalization of this relation characterizes how generally effective stellar
feedback is at driving galactic winds and outflows.  The index $\alpha$ is a
measure of how this effectiveness decreases for larger halos, and ultimately
results in a shutdown of large-scale outflows at high enough mass.  This makes
this scaling relation a key parameter to most semi-analytic galaxy formation
models (e.g \citet{Cole2000}), and an attractive target for both theorists and
observers alike.  Two primary modes for driving galactic winds have been
proposed.  Energy-driven winds (such as those investigated by
\citet{MacLow1988,Tegmark1993}, etc.) assume the cooling times for outflowing
material are shorter than the time required for a superbubble to break out of
the galactic disc.  These adiabatically expanding bubbles result in galactic
winds with mass loadings that scale as $\eta \propto v_c^{-2}$.  A recent study
by \citet{Christensen2015} showed that SN driven winds follow this scaling,
using mock observations of a sample of 7 simulated galaxies.  If the cooling
times are instead much shorter than the break out time, then outflows become
momentum driven \citep{Murray2005}, and the mass loadings instead are expected
to scale as $\eta \propto v_c^{-1}$.  \citet{Peeples2011} used a semi-empirical
model to fit the mass-metallicity relationship in low redshift galaxies to
predict that the mass loading index must be $\alpha=3$ or steeper in order to
produce a mass-metallicity relation as steep as is observed.

As in \citet{Keller2015}, we define our outflow rate using the
simple formula below, for gas particles that have $v_r > 0$, and are found
between $0.1R_{vir}$ and $R_{vir}$:
\begin{equation}
    \dot M_{out} = \sum_{r_i\in\mathrm{shell}}\frac{M_i \vec v_i\cdot\hat r}{0.9R_{vir}}
    \label{halo_massflux}
\end{equation}
This choice reflects the nature of these outflows, where gas at high
temperatures and low densities drifts outwards from the galaxy towards
the virial radius.  We chose a relatively thick shell to avoid the need for
frequent simulation outputs.  As such, our ability to resolve short-term bursts
is somewhat reduced, but the overall outflow rates trace those from a test case
with $8\times$ more frequent outputs, and a shell $8\times$ thinner.  An
excellent discussion of how to calculate outflow rates can be found in Appendix
A of \citet{Muratov2015}.

If we take the mean mass loadings $\eta = \dot M_{out}/\rm{SFR}$ of our two
populations, we see in figure~\ref{massloading_time} that the mass loading for
the unregulated population drops below $\eta=1$ at $z\sim0.5$, while the
well-regulated population never drops below $\eta\sim2$.  This helps explain the
correlation between the central concentration and stellar mass fraction (in
particular, the dichotomy between the two populations, with low central baryon
masses for well-regulated galaxies and high central masses for the unregulated
galaxies).  Failure to eject gas through outflows gives a higher central baryon
mass, and those baryons inevitably become stars.  However, we know that the
unregulated population is on average heavier (both the full halo, as well as the
central baryons).  Does this earlier drop in outflow mass loading come about
simply because the unregulated population gets heavier earlier?

\begin{figure}
    \includegraphics[width=\columnwidth]{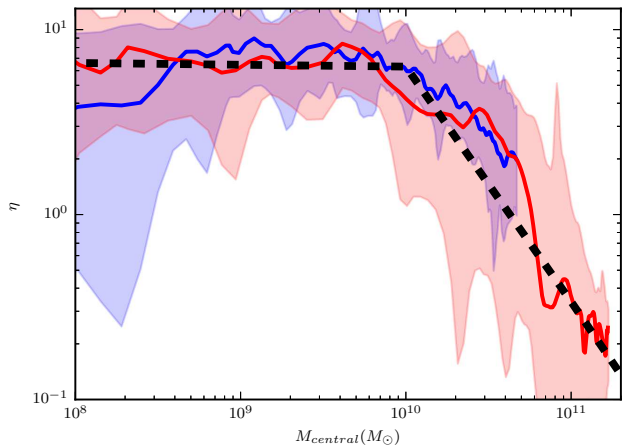}
    \caption{The mass loadings of the two populations are even more in agreement
    if we look at their relation to the central baryonic mass.  As with the mass
    loading vs. halo mass relation, we can fit a broken power law to this relation
    (shown here as the dashed line).}
    \label{massloading_central}
\end{figure}

Figure~\ref{massloading_halo} suggests exactly that.  The mass loadings appear
roughly constant, at $\eta\sim10$ for most of the mass-range for both
populations, but begin to fall at essentially the same mass, at what appears to
be the same rate.  An even tighter relation is seen in
figure~\ref{massloading_central}, when the relation between the central baryon
mass, rather than just the halo mass, is examined. It appears that for the
$\eta-M_{halo}$ and $\eta-M_{central}$ relation, a broken power-law fit, as
defined below, describes both the regulated and unregulated population.
\begin{equation}
    \eta = 
    \begin{cases}
        \alpha M^\beta  & \quad \text{if } M > M_0 \\
        (\alpha M_0^{\beta-\gamma}) M^\gamma & \quad \text{if } M < M_0 \\
    \end{cases}
    \label{outflow_scaling}
\end{equation}

Using a simple non-linear least squares fit in the on the parameters $\alpha$,
$\beta$, $\gamma$, and $M_0$, we find that the most likely value for these
parameters are $\alpha=0.9\pm0.5$, $\beta=-0.01\pm0.06$, $\gamma=-1.3\pm0.1$,
$M_{0}=10^{10.0\pm0.1}$ for the $\eta-M_{central}$ relation, and $\alpha=1\pm1$,
$\beta=0.0\pm0.1$, $\gamma=-1.8\pm0.2$, $M_{0}=10^{11.37\pm0.08}$ for the
$\eta-M_{halo}$ relation.  A fit for both of the populations independently is
consistent with these values derived for the full set of MUGS2 galaxies.
Interestingly, {\it neither} of these slopes is consistent with simple energy
or momentum driven winds (although a single power-law fit does give a slope of
$\sim0.5\pm0.2$, consistent with an energy-driven scenario, albeit with a much
weaker fit).  These values are consistent, however, with the constraints derived
for the mass-metallicity relation and gas content in nearby galaxies derived by
\citet{Peeples2011} (namely, that $\alpha$ in the $\eta\propto v_c^{-\alpha}$
relation be steeper than 3).

\begin{figure}
    \includegraphics[width=\columnwidth]{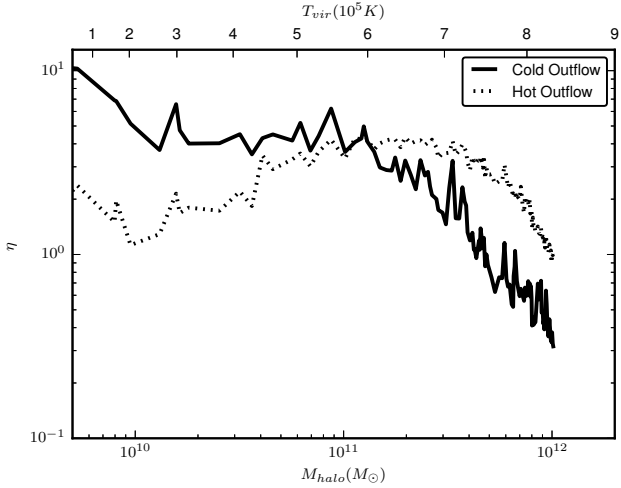}
    \caption{Mean outflow mass loading for the full sample, as a function of
    halo mass/virial temperature, and split into cold ($T<10^5\K$) and hot
    ($T>10^5\K$) components.  For low-mass halos, outflows are dominated by
    relatively cold gas (cooler superbubbles, and entrained material), while heavier
    halos have primarily hot outflows.  The transition occurs at a halo mass of
    $M_{halo}\sim10^{11}\;\Msun$, or a virial temperature of $6\times10^5\K$.}
    \label{coldmassloading}
\end{figure}

The temperature of the outflows also shows an interesting
trend as a function of halo mass.  Figure~\ref{coldmassloading} shows that as
halo mass increases, the mass loading of outflows with cold gas ($T<10^5\K$.
We choose this as a cutoff point as it lies near the peak of the cooling curve,
where we would expect, as \citet{Woods2014} showed, very little gas to be found)
monotonically decreases, while the loading of hot gas is convex, peaking at
$M_{halo} = 2-5\times10^{11}\;\Msun$, and then rapidly falling as halo mass
increases.  If we think in terms of virial temperature, this makes sense.  At
low $T_{vir}$, gas around $10^5\K$ is buoyant, and will be driven out of the
disc.  The halo itself is smaller, with a shallower potential well, thus making
it easier for the same amount of feedback energy to eject a larger amount of
entrained gas.  As the halo mass grows, eventually only the hottest gas is able
to escape, and only if it isn't weighed down by too much entrained material.
Eventually, this effect means that the total mass loading begins to drop
significantly.

\begin{figure}
    \includegraphics[width=\columnwidth]{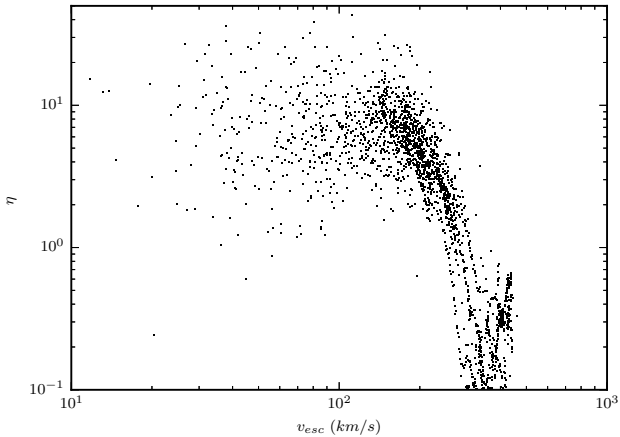}
    \caption{Mean outflow mass loading for the full sample, as a function of the
        escape velocity on the inner surface of our outflow shell
        ($0.1R_{vir}$).  Each galaxy is sampled 128 times over its evolution,
        giving us 2304 points across a range of redshifts.  As can be seen, at
        $v_{esc}\sim 250\;km/s$, the mass loading begins to fall precipitously.
        This corresponds to the sound speed of solar-metallicity gas at a
        temperature of $3.5\times10^{6}\K$.  This means that for gas at or
        above this temperature, sound waves alone are able to propel material
        significantly above the disc, and that kinetic energy losses due to
        shocks will be minimal.  Gas cooler than that will need to be propelled
        supersonically, and will drive shocks in the CGM.}
    \label{massloading_escape}
\end{figure}

We can see clearly in figure~\ref{massloading_escape} that our winds essentially
have a maximum effective velocity of $\sim250\;km/s$.  When the escape velocity
at the edge of the galaxy (where we begin to measure outflows, at $0.1R_{vir}$)
exceeds this velocity, outflow mass loadings precipitously drop from $\sim10$ to
$<1$.  As \citet{Keller2014} showed, superbubble-heated gas tends to reach an
equilibrium temperature of a a few million K, meaning that the
outflow rates appear to fall significantly when outflowing gas, moving faster
than the escape velocity from the disc is moving supersonically relative to the
superbubble-heated gas.  Radiative losses from cooling shocks may be important
here for causing the drop in outflow efficiency that produces our two
populations.


\section{Discussion}
In \citet{Keller2015}, we showed that galactic outflows, driven by
pure SN feedback modeled using the superbubble method of \citet{Keller2014} can
produce a moderate mass  $L*$ galaxy with a realistic star formation history and
no significant bulge component.  We have extended this work to a sample
of 18 galaxies, covering a mass range of roughly an order of magnitude around
that first galaxy.  This allows us to probe the expected transition
region, where SN feedback begins to fail as a self-regulation mechanism.

The transition between SN regulated galaxies at low mass and AGN regulated galaxies
at high mass has been studied observationally in the past decade.  \citet{Shankar2006}
found that the relation between halo mass and stellar mass, $r^*-\rm{band}$
luminosity, or black hole mass all were characterized by double power laws, with
breaks at $M_{vir}\sim3\times10^{11}\;\Msun$ or $M_*\sim1\times10^{10}\;\Msun$.
They interpreted these results as the transition between the scalings in SN and
AGN regulated galaxies, and constructed a simple analytic model to show the
plausibility of this idea.  \citet{Croton2006} showed that the most massive
galaxies could be produced in a semi-analytic model only when feedback
luminosity no longer simply followed the star formation rate (as it would in
stellar feedback), but instead scaled with an estimated black-hole accretion
lumininosity (as one would expect from AGN feedback).  

The most striking feature of this sample of simulated galaxies is the sharp
divide between the two identified populations: the well-regulated galaxies and
the unregulated galaxies.  The well-regulated galaxies are bulgeless and blue,
with a surfeit of extra-planar HI gas.  The unregulated sample are red and
bulge-dominated, with steep rotation curves, with maximum circular velocities
as high as $700\;km/s$.  The members of the well-regulated population have stellar mass
fractions below $4\%$, while unregulated galaxies have stellar masses all
above $6\%$, with some approaching the cosmic baryon fraction.  Clearly, the
unregulated galaxies do not match the observed properties of $L*$ galaxies in
the nearby universe.  Supernovae feedback in these galaxies has failed to
prevent runaway bulge growth \& star formation.  It has failed because it has
failed to efficiently drive winds, which \citet{Keller2015} showed to be essential for
the formation of a realistic $L*$ galaxy.  

Figure~\ref{SMHMR} shows that while the unregulated galaxies tend to be
more massive than the well-regulated ones, there is some overlap in the
$M_{halo}=8-12\times10^{12}\;\Msun$ range, where there are both unregulated
and well-regulated galaxies.  Figure~\ref{stellar_fraction_time} and
figure~\ref{SMHMR_time} show that the process of runaway star formation
happens rapidly, and can begins over a wide range of halo masses.  This is
further evidence that something beyond simply the mass of the halo that hosts a
galaxy is at play in determining whether or not supernova fail to regulate the
growth of that galaxy.

A much tighter correlation is seen when looking at the relation between the
central baryon mass (essentially, the mass of baryons in the galaxy disc), and
the stellar mass fraction.  Figures~\ref{stellar_central}
and~\ref{stellar_central_time} show that the stellar mass and central mass are
tightly correlated, and that once the central mass reaches $\sim10^{11}\;\Msun$,
SN feedback will begin to fail.  This is not particularly surprising, as stars
form in the disc, and if SFR occurred at a constant global efficiency, we would
expect to see a correlation between the central baryon mass and the stellar
mass.  However, the two populations show quite different mean star formation
efficiencies, with only $\sim38\%$ of the well-regulated disc baryons existing
in stars compared to $\sim70\%$ in the unregulated population.  While propelling
gas beyond the halo's virial radius will certainly ensure that material does not
form stars or collapse through the disc to build a bulge, removing material from
the halo isn't necessary to preventing runaway star formation/bulge growth.
Material cycling through the CGM, propelled to a few $100\;\kpc$ above the disc,
can take hundreds of Myr to re-accrete back onto the disc.  This means that the
increasing escape velocity of the disc is more important to preventing outflows
from regulating star formation than an increase in the total halo mass.  A more
massive halo, with a relatively light disc can still have high mass loadings in
it's SN-driven outflows, and still have those outflows remove star forming
material from the disc for much a galaxy's life.

\subsection{Outflow Scaling Relations and The End of Regulation}
For gas ejected by stellar feedback, the path out of a galaxy disc is fraught.
To actually escape the disc, buoyant gas must push through the gas above it.
High resolution simulations of disc slices have shown that this means SN
occurring higher above or below the disc drive hotter, faster outflows
\citep{vonGlasow2013,Sarkar2015}.  These results have tended to show that when
feedback is deposited randomly, as opposed to at density peaks, stronger
outflows can be driven.  This is problematic for the simple fact that star
formation should be occurring most vigorously in the densest ISM gas, and this
should therefore be the site of most feedback deposition.  Drift of stars from
their natal environment can give some offset between dense gas and where massive
stars ultimately end up, but it is unlikely that this is a significant effect
for the majority of stars.  \citet{Governato2010,vonGlasow2013,Christensen2015}
showed that when SN feedback is clustered, resulting in stronger outflows
compared to the same number of SN spread
distributed more smoothly throughout the disc.  Unfortunately, even these
high-resolution, well controlled studies have found significant variation in how
wind mass loadings scale with halo mass.  The relationship between circular
velocity and mass loading has been found to vary both as a function of how
feedback energy is injected \citep{vonGlasow2013} and what scaling relationship
was used to determine halo mass from of the slice (\citet{Creasey2013} found
that using different scaling relations between the halo mass and surface density
could give an index $\alpha$ that varies from 2.5 to 4.8).  These fully cosmological
simulations have the advantage of removing this second source of uncertainty.

The failure of SN to regulate the more massive, disc-heavy galaxies is
ultimately a consequence of the relationship between the efficiency of SN
powered outflows and the mass of the halo (or more precisely, the disc).
Figures~\ref{massloading_halo} and~\ref{massloading_central} show that the
outflow mass loading is characterized by a broken power-law, with roughly
constant mass loadings at low mass, and mass loadings that follow a power-law
with negative slope once a critical mass is exceeded.  This model for outflow
mass loadings is similar to one explored by \citet{Font2011}.  They found that,
in a semi-analytic study of the Milky Way satellites, the best fit for the
observed luminosity function was a so-called ``saturated feedback'' model, where
mass loadings were flat at low masses, and decreased as a constant power-law in
halos with $v_c>65\;km/s$ (somewhat lower than the critical value we have
found).  Further evidence against a single power-law in the $\eta-M$ relation
came from recent work by \citet{Hou2015}.  By combining the analysis of
\citet{Font2011} along with more recent MW satellite data, the mass-metallicity
relation in local galaxies, and estimates of the redshift of reionization, they
found that only a broken power-law for the $\eta-M$ relation can fit all of the
observed data.  Not only that, their ``saturated'' model resembles quite closely
the fit we have found here, with a flat slope at low masses, and $\eta\propto
M_{halo}^{-1.1}$ for high masses.
\citep{Muratov2015} found a nearly flat $\eta-M_{halo}$ relation
using the FIRE feedback model, with $\eta\propto M_{halo}^{-0.35}$.

This relation seems to arise as a result of different $\eta-M$ relations for
cold and hot gas.  The combination of a steadily decreasing mass loading for
cold gas, combined with hot gas mass loadings that peak at
$M_{halo}\sim5\times10^{11}\;\Msun$ ultimately result in the broken power-law
we see for all outflowing material.  In an upcoming study, we will be examining
how this relation ultimately arises through detailed examination of ultra high
resolution simulations of outflow regions.  A detailed study of the
hydrodynamics of wind venting should determine the processes by which cold ISM
is entrained within the hot outflowing gas, and hopefully explain the origin of
the $\eta-M$ relations we have found here.

The results here also show that even in the unregulated population, superbubbles
are still able to break out of the galaxy disc (as is shown by the adiabatic
evolution of feedback-heated gas seen in figure~\ref{phase}).  In order for a superbubble to cool
adiabatically by a factor of $\sim100$, it must be able to expand by that same
factor, and only once it has left the galaxy disc is there enough room for it to
do this.  The failure of SN feedback in the massive, unregulated population is
an issue of a drop in efficiency, rather than a complete shutoff, of galactic
outflows.  The reason the unregulated population exists is that the galaxies in
this population spent roughly half of their lifetime above the critical mass in
the $\eta-M$ relation, where SN cannot efficiently drive outflows.  This has
resulted in these galaxies becoming too centrally concentrated, and vastly
overproducing stars. Figure~\ref{coldmassloading} and
figure~\ref{massloading_escape} show that at higher masses (either for the halo
itself, or simply for the interior regions), only the hottest material is able
to escape from the disc.  

The $\sim250\;km/s$ break in mass loading seen in
figure~\ref{massloading_escape} corresponds to the sound speed of relatively hot
gas ($3.5\times10^6\K$ for solar metallicity).  Even with no radiative losses
whatsoever, if $10^{51}\;erg$ is deposited into $1000\;\Msun$ of gas (the specific
energy that $\eta=10$ would  yield), this gas would have a temperature of
$2.5\times10^7\K$, and a sound speed of only $\sim210\;km/s$.  Thus, it is
somewhat unsurprising that the mass-loadings seen here only stay high for escape
velocities below $200\;km/s$.  With mass-loadings falling to unity at escape
velocities of almost exactly $300\;km/s$, we can infer that the total losses
(radiatively and otherwise) between the initial SN explosions and the subsequent
outflows are $\sim80\%$.  The roughly constant outflow rates seen at lower
masses/escape velocities imply that a process other than radiative losses is
limiting the outflow mass-loadings.  The hydrodynamic coupling between hot,
outflowing gas and the surrounding medium (the key process that is involved in
setting these mass loadings) will be the subject of a subsequent study.

\subsection{Runaway Bulge Growth points to AGN Feedback}

The existence of an unregulated population of galaxies raises the
question of what physical mechanism is missing in these simulations.  What is
important to consider is that this missing physics must be more effective in the
high-mass, bulge dominated unregulated population without causing the low
mass/well-regulated population to underproduce stars, or disrupt their discs.

While cosmic rays may be important in certain ISM environments, and may drive
the fastest components of galactic winds, there is no known mechanism involved
in CR feedback that suggests they become more effective at higher mass.  In
particular, there is nothing to suggest that the critical mass we see at
$M_{halo}\sim10^{12}\Msun$ is in any way related to the efficiency or
effectiveness of CR feedback.  For the unregulated population, previous results
suggest that cosmic rays alone will not be able to produce outflows any more
effectively than we have seen SN alone drive, as the mass-loadings they can
produce are typically $\ll10$.  Past simulations have also shown that for the
mass range where SN feedback begins to fail to regulate star formation,
radiation pressure either fails to prevent overproduction of stars
\citep{Aumer2013}, or completely disrupts the thin disc \citep{Roskar2014},
producing ``puffy'', spherical galaxies with large stellar scale heights,
depending on the details of the subgrid model and how that translates into a
coupling between the photon fluid and the ISM.

The fact that SN begin to fail at relatively high masses hints that the
mechanism that begins to take over in regulating the growth of the galaxy is
AGN.  It has long been suspected that the high-mass end of galaxy evolution was
dominated by AGN feedback from the growth of their SMBHs, and this has been
explored heavily in semi-analytic models
\citep{Bower2006,Croton2006,Shankar2006}.  A survey of SDSS galaxies with found
that the detectable AGN fraction was essentially zero below
$M_*\sim10^{10}\;\Msun$, rising to $\sim50\%$ by
$M_*\sim5\times10^{10}\;\Msun$, with essentially all galaxies with
$M_*>5\times10^{11}\;\Msun$ hosting an AGN \citep{Kauffmann2003}.  This
transition mass lies exactly where our two populations begin to diverge.  No
galaxy in our sample with $M_*<4\times10^{10}\;\Msun$ is unregulated, and all
but two of the unregulated galaxies have stellar masses above
$10^{11}\;\Msun$.  This is particularly interesting when considering the
population of ``powerful'' AGN in the SDSS peaks at $M_*\sim10^{11}\;\Msun$.
If these powerful AGN effectively quench star formation within their host
galaxies, clamping the stellar mass at or below $10^{11}\Msun$, this mechanism
would put essentially all of our unregulated population back within the
\citet{Hudson2015} SMHMR.  If these AGN also ejected a significant fraction of
the gas within the inner $<1\kpc$ of the unregulated population, it would also
solve the problem of their unrealistically large bulges \& peaked rotation
curves.

The correlation between black hole mass and bulge mass (see \citet{Kormendy2013}
for a review of this evidence) is strong evidence that the central regions
co-evolve with with the galaxy (pseudo-)bulge (to first order, this of course
must be the case: they are formed from the same material, in roughly the same
place).  The fact that our unregulated population has grown a much too
massive bulge makes AGN feedback an attractive potential resolution.  In order
to build these massive bulges, angular momentum losses in the gas disc must have
been significant, funneling material down towards the center of the disc.  This
naturally would be a necessary component to fuelling the growth of a SMBH, and
powering an AGN.  The well-regulated population likely would
under produce stars if AGN feedback was a strong effect in those galaxies.  
Conveniently, the lack of any sort of significant bulge component means that the
SMBH within those galaxies would likely be starved of fuel, preventing any
significant, sustained AGN feedback.  Further evidence for the negligible effect
of AGN feedback for these smaller galaxies comes again from the SDSS, where it
can be seen that galaxies with $M_*<5\times10^{10}\;\Msun$ contain a powerful
AGN in less than $5\%$ of the cases.

\section{Conclusion}
By correctly modelling supernova-driven superbubble feedback, we allow
clustered star formation to efficiently drive outflows.  This allows SN
feedback to produce lower mass galaxies with flat rotation curves, realistic
SMHMRs, and small bulges.  SN feedback breaks down as a regulator of stellar
mass and bulge growth in galaxies with halo masses $>10^{12}\Msun$ (or stellar
masses > $4 \times 10^{10}\Msun$).  This breakdown produces a distinct pair of
populations.   Galaxies below these
critical masses have stellar masses within the observed SMHMR, flat rotation
curves, and more mass loaded winds over their evolutionary history.
When simulated with SN feedback alone,  galaxies
above this mass have stellar fractions approaching the cosmic baryon fraction,
redder stellar populations, steeply peaked rotation curves with maximum circular
velocities as high as $700\;km/s$, and relatively inefficient outflows.  As
massive galaxies grow, they move from the well-regulated to unregulated
population quickly, typically producing the bulk of their stars in a burst
lasting $\le1\;Gyr$, after which star formation slows simply due to a dearth of
gas to form stars out of.

The cause for this transition is the relation between the mass loading
of superbubble outflows $\eta$ and the halo or disc mass.  A critical value
exists for both of these masses, where mass loadings begin to drop steeply as
mass increases, reducing the ability of SN to regulate the SFR and baryon
content of the galaxy.  Once the escape velocity from the inner $0.1R_{vir}$
exceeds $\sim250\;\kms$, mass loadings fall rapidly to $>1$.  A simple broken
power-law fit describes the relation between the outflow mass loading and both
the halo mass and disc mass.  For disc mass below $10^{10}\;\Msun$ and
halo mass below $2 \times 10^{11}\;\Msun$, outflow mass loadings are
approximately constant, with $\eta\sim8$.
        
SN feedback begins to fail at exactly the mass range that strong AGN are
observationally detected.  This coincides with the runaway growth of massive
stellar bulges in the unregulated population.  The associated black hole feeding
and nuclear activity should not only regulate bulge growth but also the total baryonic
distribution and star formation for galaxies at the high mass end.  
Up to this mass, supernova-driven superbubbles alone can regulate the
baryonic content and star formation in galaxies.

\section*{Acknowledgements}
The analysis was performed using the pynbody package
(\texttt{http://pynbody.github.io/}, \citep{pynbody}).  We thank Alyson Brooks,
Fabio Governato, Tom Quinn, and John Kormendy for useful conversations regarding
this paper.  The simulations were performed on the clusters hosted on
\textsc{scinet}, part of ComputeCanada.  We greatly appreciate the contributions
of these computing allocations.  We also thank NSERC for funding supporting this
research.
\bibliographystyle{mnras}
\bibliography{references}

\clearpage
\end{document}